\documentclass[usegraphicx,usenatbib]{mn2e}
\usepackage{subfigure}
\usepackage{longtable}
\usepackage{graphicx}
\usepackage{graphics}
\usepackage{keyval}
\usepackage{trig}
\usepackage{graphicx}% Include figure files
\usepackage{dcolumn}% Align table columns on decimal point
\usepackage{bm}% bold math
\def\beq{\begin{equation}}
\def\eeq{\end{equation}}
\def\bey{\begin{eqnarray}}
\def\eey{\end{eqnarray}}
\def\msun{M_\odot}
\def\lsun{L_\odot}
\def\kms{\, {\rm km \, s}^{-1} }

\def\prd{Phys. Rev. D}
\def\mnras{MNRAS}
\def\apj{ApJ}

\def\apjs{ApJ}
\def\apjl{ApJ}
\def\araa{ARAA}
	
\def\aap{A \& A}
\def\aj{AJ}
\def\lcdm{{\Lambda}CDM}
\def\half{{1 \over 2}}
\def\numb{30}
\def\aap{Astron. Astrophys.}
\begin{document}

\title{Equilibrium configurations of $11~eV$ sterile neutrinos in MONDian galaxy clusters}

\author[G. W. Angus, B. Famaey, A. Diaferio]{G. W. Angus$^{1,2}$\footnote{email:angus@ph.unito.it}, B. Famaey$^{3,4}$, A. Diaferio$^{1,2}$ \\
$^{1}$Dipartimento di Fisica Generale ``Amedeo Avogadro", Universit\`a degli studi di Torino, Via P. Giuria 1, I-10125, Torino, Italy \\
$^{2}$Istituto Nazionale di Fisica Nucleare (INFN), Sezione di Torino, Torino, Italy\\
$^{3}$AIfA, Universit\"at Bonn, 53121 Bonn, Germany\\
$^{4}$Observatoire Astronomique, Universit\'e de Strasbourg, CNRS UMR 7550, F-67000 Strasbourg, France\\}

\date{\today}
\maketitle
\begin{abstract}
Modified Newtonian Dynamics (MOND) can fit a broad range of galaxy kinematic data, but struggles with clusters of galaxies. MONDian clusters need dark matter, and here we test the $11~eV/c^2$ sterile neutrino - used to fit the first three acoustic peaks of the cosmic microwave background - by investigating their equilibrium distributions in $\numb$ groups and clusters over a wide range of temperatures. We do this by first taking the known sterile neutrino density, necessary for hydrostatic equilibrium of the intracluster medium (or to produce the observed lensing map). Then, we solve for the sterile neutrino velocity dispersion, needed for their own hydrostatic equilibrium, through the equation of state for a partially degenerate neutrino gas. The velocity dispersion is a unique, continuous function of radius determined by the density and mass of the sterile neutrinos particles. Knowing both the sterile neutrino density and velocity dispersion tells us the Tremaine-Gunn phase-space limit at all radii. We find that all $\numb$ systems serendipitously reach the Tremaine-Gunn limit by the centre, which means a portion of the dynamical mass must always be covered by the brightest cluster galaxy. Interestingly, the typical fitted K-band mass-to-light ratio is unity and at most 1.2, which is very consistent - although leaving no margin for error - with stellar population synthesis models.  Amidst the sample there are several special cases including the Coma cluster (for which dark matter was first proposed),  NGC~720 (where geometrical evidence for dark matter was found) and the bullet cluster (where dark matter - of some kind - in clusters was directly proven to exist). We demonstrate that $11~eV/c^2$ sterile neutrinos are unlikely to influence spiral galaxy rotation curves, as they don't influence even some very massive early-types (NGC~4125 and NGC~6482). Finally, we conclude that it is intriguing that the minimum mass of sterile neutrino particle that can match the cosmic microwave background is identical to the minimum mass found here to be consistent with equilibrium configurations of MONDian clusters of galaxies. 
\end{abstract}

\section{Introduction}
\protect\label{sec:intr}
Sterile neutrinos are hypothetical additions to the standard model of particle physics. They are right handed, neutral leptons which interact only via gravity, which earns them the ``sterile'' prefix, contrary to the active neutrinos which also participate in the weak interaction. Gravity aside, sterile neutrinos also can interact with the active neutrinos via the quantum mechanical phenomenon of neutrino oscillations. This behaviour has been investigated by the Liquid Scintillator Neutrino Detector (\citealt{aguilar01}) and the MiniBoone experiment (\citealt{maltoni07}). However, no concrete evidence was convincingly found to suggest the existence, or non-existence, of sterile neutrinos from the disappearance of active neutrinos. The basis of some appeals to sterile neutrinos are for aesthetic reasons, since the active neutrinos are entirely left-handed; and others make use of them in the so called ``see-saw mechanism'' which can give rise to the small masses of the active neutrinos (\citealt{lindner02}). Here we make no claims of a deeper theory for sterile neutrinos, but rather continue to investigate a startling coincidence.

A recent analysis of the Cosmic Microwave Background (CMB) by \cite{angus09} has demonstrated that the acoustic peaks in the angular power spectrum as measured by WMAP (\citealt{dunkley09}) and ACBAR (\citealt{reichardt09}) can convincingly be generated by a single, thermal (by virtue of neutrino oscillations in the early Universe) sterile neutrino (SN) with mass $m_{\nu_s}=11~eV/c^2$. This SN is a straight substitution for the cold dark matter (CDM) of the concordance cosmological model (\citealt{spergel07}), such that $\Omega_{\nu_s}=0.0205m_{\nu_s}=0.225$, $\Omega_{b}=0.047$ and the spectral index $n_s=0.965$. However, whereas the unknown CDM particle can condense and form structures on virtually any scale, which puts it at odds with certain observations - like the lack of substructure as compared to the cosmological N-body simulations (\citealt{klypin99,moore99,kroupa05}) and the existence of DM cores in galaxy halos (\citealt{gnedin02,gentile04}) - the hot SN free streams out of galaxies. This makes it a non-starter as a traditional dark matter candidate, since it would render spiral galaxies (like the Milky Way) and dwarf galaxies (like Draco and Ursa Minor) incapable of retaining their stars.

Fortunately, there is a well studied alternative to the standard lore of gravity called Modified Newtonian Dynamics (or MOND; see \citealt{milgrom83a,sanders02,bekenstein06,milgrom08} for reviews) which has an uncanny knack of predicting the dynamics of galaxies from a simple relation using only the baryonic matter density. The paradigm of MOND introduces a new physical constant with dimensions of acceleration, $a_o=1.2\times10^{-10}ms^{-2}$, around and below which dynamics do not follow from standard Newtonian theory. In particular, the true modulus of gravity, $g(r)=V_c(r)^2 r^{-1}$, is not linearly related to the Newtonian gravity, $g_n(r)=GM(r)r^{-2}$, but instead 

\beq
%g = g_N \frac{\left(1+\sqrt{1+4a_o^2/g_N^2}\right)^{1/2}}{\sqrt{2}}
g = {\half} g_n  \left[1 + \sqrt{1+{4a_o \over g_n}}\right]
\eeq
%${(g/a_o) \over \sqrt{1+(g/a_o)^2}}g=g_n$. 
This ensures adherence to two critical axioms: that gravity is Newtonian in regions of strong gravity, and that when $g << a_o$, $g \propto 1/r$ meaning rotation curves are flat at the periphery of spiral galaxies.

The additional gravity afforded by MOND, replaces the need for DM in dwarf spheroidals (\citealt{angus08,milgrom95,sanchez07,serra09}), spiral (e.g. \citealt{mcgaughdeblok,sandnoord,fb05,mcgaugh08}) and X-ray dim elliptical galaxies (\citealt{milgrom03,aftcz}) often with remarkable accuracy. Furthermore, rotation curves of tidal dwarf galaxies have been observed by \cite{bournaud07} and the independent analysis by \cite{gentile07a} and \cite{milgrom07b} reveal not only their consistency with MOND (with zero adjusted parameters), but also constitutes a direct falsification of dark matter being made only of CDM in galaxies: assuming the data are reliable. What is more, \cite{kroupa05} have advocated that the Milky Way's dwarf spheroidal galaxies are also tidal dwarf galaxies. This convincing demonstration of MOND's predictive ability does not, however, extend to the realms of clusters of galaxies.

An obvious difference between galaxies and clusters of galaxies is simply the scale involved. If a typical galaxy is one or two tens of $kpc$ across, clusters have accurate measurements of the gravitational potential (through the intracluster medium or weak gravitational lensing) on scales ten times that and therefore volumes many thousand times greater. Studies of the dynamics of groups and clusters in MOND (e.g. \citealt{the88,sanders94,sanders99,aguirre01,sanders03,point,aszf,sanders07,feix08,milgrom08}) have all shown that there is a huge central mass deficit and \cite{afb} demonstrated that active neutrinos, even at the experimental maximum ($\sim 2~eV/c^2$), cannot clump densely enough. A more plausible solution is the $11~eV/c^2$ SN proposed by \cite{angus09} to fit the first three peaks of the CMB.

The feasible particle mass of SN is fully determined by fitting the CMB power spectrum, and there is virtually no freedom available above 5\% of the $11~eV/c^2$ mass. It is the particle mass, and particle mass alone, that sets the properties of the SNs in clusters and galaxies. Therefore, it is a highly intriguing corollary that the mass required to match the CMB is in the tiny range of neutrino masses (perhaps $11-12~eV/c^2$) that can both free stream out of galaxies (MOND does not require any DM in galaxies) and clump densely enough in galaxy clusters to account for the serious mass deficit exposed there.

We now have the basis of a predictive cosmological model, where we have made two positive trades. Firstly we exchange Newtonian dynamics for Modified Newtonian Dynamics, which helps explain in detail the origin of the mass discrepancy in all galaxies, and with fewer freedoms than the often contrived CDM halos (as demonstrated by \citealt{mcgaugh05b,kdn}). Secondly, we swap CDM for an $11~eV/c^2$ SN. The added bonus of SNs over CDM, is that knowing only the particle mass gives fixed predictions for the CMB and for structure formation (with a small dependence on the $\mu$-function in only the latter case).

In the very early Universe, neutrino decoupling (both active and sterile) occurs at a temperature of $kT \sim 1~MeV$. This means that the $11~eV/c^2$ SNs (as well as the sub $eV/c^2$ active neutrinos) are ultra-relativistic during decoupling, which freezes in their Fermi-Dirac distribution and their cosmological abundance is fixed (e.g. \citealt{peacock}). Typical CDM candidates, like 100~$GeV$ neutralinos (e.g. \citealt{hofmann01}) or $>100~keV/c^2$ SNs (see the review by \citealt{boyarksy09}) are non-relativistic during decoupling, and self-annihilations must be used to tune the cosmological abundance. This applies to all CDM candidates which means knowing the mass of a CDM particle tells us the mass and not the cosmological abundance. Free parameters, like the interaction cross-section, must explain why they have the correct cosmological abundance.

In addition, while searching for an explanation of the anomalous low energy excess of electron neutrinos observed by the Miniboone experiment, \cite{giunti08} found agreement with the data by postulating a perfectly plausible renormalisation of the original flux of muon-neutrinos and oscillations (that are energy dependent) from electron neutrinos to $11\pm7~eV/c^2$ SNs. This hypothesis clearly warrants further investigation, and will be testable with the T2K neutrino experiment (\citealt{hastings09}) which will have a near detector at $L=280~m$ giving it a similar L/E (E being neutrino energy) as MiniBoone (M. Laveder, private communication).

In this paper, we seek to investigate the $11~eV/c^2$ SN's influence in clusters of galaxies in MOND. Primarily, we need to ascertain that every cluster has an equilibrium distribution of neutrinos that can account for all the gravitating mass. This is by no means guaranteed since there is a maximum density set by the Tremaine-Gunn limit (\citealt{tremaine79}; see \S\ref{sec:psc}) depending on the velocity dispersion of neutrinos and the mass of an individual particle. If the required density of the equilibrium models exceed the Tremaine-Gunn limit (or if the required $M/L$ of the BCG to keep the SN density below the Tremaine-Gunn limit is too high), they would be excluded. To this end we have taken a relatively large sample of $\numb$ relaxed galaxy groups and clusters to gauge their consistency with the missing mass problem of clusters of galaxies in MOND.

\section{Data}
We take the temperature and density profiles of the intracluster medium (ICM) in 7 clusters of galaxies with temperatures in the range of $1.8-9~keV$ from the sample of \cite{vikhlinin06}, a further 13 groups of galaxies from the sample of \cite{gastaldello07}, the cluster A~2589 from \cite{zappacosta06} and 3 X-ray bright, isolated early type galaxies from the sample of \cite{humphrey06}. This covers the full sample used in \cite{afb}, but in addition, we make an analysis of the two clusters comprising the bullet cluster (\citealt{clowe04,clowe06,bradac06,aszf}) working from the NFW halos fitted to the weak-lensing convergence map. We also take a fitted NFW profile for the cluster A~1689 from the strong lensing analysis of \cite{halkola06}, take 3 estimates for the density profile of the Coma cluster (\citealt{gavazzi09,kubo07}), a pair of clusters in the Lynx field (\citealt{jee06}) and the fossil group RXJ~1416 (\citealt{khosroshahi06} ).

 For the \cite{gastaldello07} sample, we only had X-ray data for 10 of the 13, so for the other 3 (A~2717, IC~1860, MS~1160) we started from their fitted NFW (\citealt{nfw97}) profiles and included the brightest cluster galaxy. 

In \cite{vikhlinin06}, the authors give fully analytical descriptions of the ICM temperature and density allowing us to solve the equation of hydrostatic equilibrium to give the dynamical mass of each cluster (see \S\ref{sec:solv}). Similarly for the groups of galaxies, we were provided with (D. Buote private communication) high resolution data that gives the ICM density and temperature as functions of radius. We fitted $\beta$-models to the ICM density and have fitted the same analytical models defined in \cite{vikhlinin06} to the temperature. These models are rather sophistocated, many parameter models that maximise the accuracy of the fits.

The masses of the brightest cluster galaxies (BCGs) are taken either from the \cite{gastaldello07,vikhlinin06,humphrey06,zappacosta06} K-band luminosity estimates, the observations of \cite{linmohr} or from K-band magnitudes on the {\it NED} database as per \cite{afb}.

Naturally, there are errors associated with the best fit density and temperature profiles, which we do not explicitly investigate here. It is enough to say that the observational uncertainties surrounding the Newtonian dynamical mass (from which everything is deduced) are no larger than the uncertainties in the triaxiality of the cluster or the interpolating function used to find the MONDian dynamical mass, so we waste little time debating them. Moreover, we use a sample containing $\numb$ of some of the most relaxed systems so that a general consensus can be reached.

\section{Solving for sterile neutrino equilibrium configurations}
\protect\label{sec:solv}
\subsection{Intracluster medium hydrostatic equilibrium} 
The density of $11~eV/c^2$ SNs (or any DM candidate) in MONDian clusters of galaxies required to provide hydrostatic equilibrium of the ICM is given by the following steps, as per \cite{afb}. Firstly, we take the observed density and temperature of the ICM, $\rho_x(r)$ and $kT_x(r)$ respectively, and numerically take their logarithmic derivative with respect to radius to find the true gravity as a function of radius
\beq
\protect\label{eqn:grav}
g(r)={-kT_x(r) \over wm_pr}\left[{d\ln \rho_x(r) \over d\ln r}+{d\ln kT_x(r) \over d\ln r}  \right],
\eeq
where $w=0.62$ is the mean molecular weight of the ICM. The gravity simply is related to the total MOND enclosed mass by
\beq
\protect\label{eqn:mm}
M_m(r)=r^2G^{-1}g(r)\mu(g/a_o),
\eeq
and the interpolating function is the simple one 
\beq
\protect\label{eqn:fb}
\mu(g/a_o)={g/a_o \over 1+g/a_o},
\eeq
(see \citealt{famaey06} for a discussion of how it fares in spirals). This single line is the only stage at which MOND is involved.

After subtracting the mass of the ICM ($M_x(r)=\int_0^r 4\pi\hat{r}^2\rho_x(\hat{r})d\hat{r}$), we are left with the SN mass distribution and the unsubtracted BCG, $M_{bcg}+M_{\nu_s}(r)=M_m(r)-1.15M_x(r)$, where the mass of the brightest cluster galaxy ($M_{bcg}=M/L_K \times L_{K,bcg}$). The 1.15, which is of virtually no consequence, multiplies the ICM mass to include the contribution of galaxies in the cluster. We invert this to give the SN density, there we ignore the BCG until later, although it is vitally important, for reasons that will become obvious later.
\beq
\protect\label{eqn:rhonu1}
\rho_{\nu_s}(r)=(4\pi r^2)^{-1}{d \over dr}M_{\nu_s}(r).
\eeq

\subsection{Sterile neutrino hydrostatic equilibrium}
At this point we have a deduced density of SNs in a cluster of given temperature profile: at least it is the density exactly required for hydrostatic equilibrium of the ICM (or in the case of lensing, to create the observed convergence map). In addition to this constraint, there is an equation for hydrostatic equilibrium of the SNs themselves, which is crucial to this analysis. This equation invokes the equation of state of neutrinos (active or sterile; see e.g. \citealt{sanders07}) and what this gives us are equations that define the density and pressure of a partially degenerate neutrino gas, coupled via the hydrostatic equilibrium relation
\beq
\protect\label{eqn:hyd}
{d \over dr}P_{\nu_s}=-\rho_{\nu_s}(r)g(r).
\eeq

To express the density, we start with the equilibrium occupation number
\beq
\protect\label{eqn:on}
f={1 \over exp\left({x-\chi}\right)+1},
\eeq
where $x$ and $\chi$ are respectively the ratios of neutrino energy and chemical potential to temperature. A large positive or negative $\chi$ denotes strong degeneracy or non-degeneracy respectively. From this we find the phase space density
\beq
\protect\label{eqn:ps}
n=g_{\nu}h^{-3}m^4_{\nu_s}f.
\eeq
Here, $h=4.136\times 10^{-15}eV.s$ is Planck's constant and $g_{\nu_s}$=2 takes into account the anti-particles, whilst remembering that neutrinos have only a solitary helicity state. One immediately sees that there is an absolute upper limit to the allowed density of SNs in phase-space, corresponding to full degeneracy ($\chi=+\infty$) and known as the Pauli limit. The starting momentum distribution of neutrinos in the early universe corresponds to half this limit.

It should be clear here that in groups and clusters of galaxies, SNs of $11~eV/c^2$ mass are non-relativistic since they are travelling at velocities of the order 100-1000$\kms$ (e.g. Fig~\ref{fig:c1} central panels). Therefore, neutrino energy is related to momentum by $E=p^2/2m_{\nu_s}$. The number density of neutrinos is given by $\int f d^3p$ and multiplying this by neutrino mass ($m_{\nu_s}=11~eV/c^2$), converting momentum to energy and temperature to velocity dispersion ($kT_{\nu_s}=m_{\nu_s}\sigma_{\nu_s}^2$) we find our equation for neutrino mass density
\beq
\protect\label{eqn:rhonu}
\rho_{\nu_s}(r)=4\sqrt{2}\pi g_{\nu} h^{-3} m_{\nu_s}^4 \sigma_{\nu_s}^3(r)F_{1/2}(\chi),
\eeq 
where the SN velocity dispersion and temperature are $\sigma_{\nu_s}$ and $kT_{\nu_s}$ respectively and $F_{1/2}(\chi)=\int_0^\infty x^{1/2}f(\chi) dx$. In a non-relativistic neutrino gas, the pressure is equal to 2/3 the internal energy per unit volume ($U_{\nu_s}=\int E(p)f d^3p$) giving

\beq
\protect\label{eqn:prenu}
P_{\nu_s}(r)={8\sqrt{2}\pi \over 3} g_{\nu} h^{-3} m_{\nu_s}^4 \sigma_{\nu_s}^5(r)F_{3/2}(\chi),
\eeq
where $F_{3/2}(\chi)=\int_0^\infty x^{3/2}f(\chi) dx$. 

It emerges that there are two variables here that must be set in order for the neutrinos to exist in hydrostatic equilibrium. Primarily, $\chi(r)$, must be fixed so that the fixed density of Eq~\ref{eqn:rhonu1} is matched by Eq~\ref{eqn:rhonu}. In case this is not obvious, we solve Eq~\ref{eqn:rhonu} for $F_{1/2}(\chi)$, for which there is a unique, continuous $\chi(r)$ once the neutrino velocity dispersion is given. There is no independent method, apart from cosmological simulations of the collapse of the baryonic plus SN two-fluid mixture to estimate the chemical potential, $\chi$, so we choose here to fit it to the data. Whether detailed numerical simulations of cluster formation in a MOND cosmology will reproduce this chemical potential will be a crucial test of the equilibrium models presented hereafter.

This $\chi(r)$ is then transfered to Eq~\ref{eqn:prenu}, and although $\chi(r)$ is used to balance $\rho_{\nu}(r)$  and ensure it remains unchanged when $\sigma_{\nu_s}$ is varied, it influences the pressure, $P_{\nu_s}$, non-linearly. Secondly, Eq~\ref{eqn:hyd} must be satisfied, however, it is not satisfied if we make the assumption that the sterile neutrino velocity dispersion (VD) is identical to the ICM VD. In general, there would be too much pressure because the neutrinos would be too hot for their given distribution.

{\it A priori}, there is no reason that the $11~eV/c^2$ SNs should have precisely the same temperature as the ICM. Both fluids are orbiting in the same gravitational potential, but there is no rapid exchange mechanism to bring them into mutual equilibrium, as there was in the very early Universe. In fact, this is likely to be related to the formation epoch of the cluster, where the SN halos would have formed at relatively high redshift while the Universe was more dense. Later, the ICM would have fallen into the deep potential well, created by the SN halo, and therefore it is logical that the ICM should be hotter than the SNs. In addition to probing the $\chi(r)$'s fitted, numerical simulations of galaxy clusters in MOND will, in time, be able to tell us if the difference in neutrino and ICM VD is realistic.

From Eqs~\ref{eqn:rhonu} \& \ref{eqn:prenu} one can see that the SN VD influences the product of the density and gravity (the right hand side of Eq~\ref{eqn:hyd}) non-linearly with respect to the gradient of the neutrino pressure. Essentially, to force the SN distribution into hydrostatic equilibrium, with its fixed density,  we can modify the SN VD. However, as we shall discover in the next section, this cannot be achieved by a simple scaling from the ICM VD, but instead by solving for the neutrino VD, $\sigma_{\nu_s}$, as a unique, continuous function at every radius.

\subsection{Solving for a unique sterile neutrino velocity dispersion}

In principle, for hydrostatic equilibrium of the SNs, we need ${{d \over dr} P_{\nu_s}(r) \over \rho_{\nu_s}(r)g(r)}=-1$ and writing this out after some algebra gives us

\beq
\protect\label{eqn:nueq1}
\sigma_{\nu_s}(r) {d \over dr}\sigma_{\nu_s}(r) +{\sigma_{\nu_s}^2(r) \over 5 F_{3/2}(r){d \over dr} F_{3/2}(r)}=-{3 \over 10}{F_{1/2}(r) \over F_{3/2}(r)}g(r).
\eeq

Using the substitution $\epsilon(r)=\sigma_{\nu_s}^2(r)$, leading to $\epsilon '(r)=2\sigma_{\nu_s}(r)\sigma_{\nu_s}'(r)$ and the integrating factor $F_{3/2}(r)^{0.4}$, we can reduce this first order linear differential equation to

\beq
\protect\label{eqn:nueq2}
\sigma_{\nu_s}^2(r)={3 \over 5} F_{3/2}(r)^{-0.4}\int_{\infty}^r {F_{1/2}(\hat{r}) \over F_{3/2}(\hat{r})^{0.6}}g(\hat{r})d\hat{r}.
\eeq

To find pressure and density in the first instance, we must submit a trial $\sigma_{\nu_s}$ to Eq~\ref{eqn:nueq2}: typically we try $\half \sigma_{x}$, where the ICM VD is defined by
\beq
\sigma_{x}=c \times \left({kT_x \over w m_p}\right)^{1/2}.
\eeq
Here, $c=3\times10^5 \kms$ is the speed of light, $kT_x$ is the ICM temperature in $keV$, $w=0.62$ is again the mean molecular weight and $m_p=9.38\times10^5~keV$ is the mass energy of a proton. From the trial solution, we iterate rapidly towards convergence. The final $\sigma_{\nu_s}(r)$ gives hydrostatic equilibrium to better than 1 per cent at all radii and most importantly is unique - set only by the SN mass and its indirectly observed  density from Eq~\ref{eqn:rhonu1}.

\subsection{Phase space constraints}
\protect\label{sec:psc}
We now have a unique correlation between the density of SNs and their velocity dispersions. This is important because these two variables define the phase space distribution of the SNs, to which there is a fundamental limit.

Liouville's theorem states that (in the absence of encounters) flow in phase space is incompressible and that each element of phase-space density is conserved along the flow lines. However, this only applies to the fine-grained phase space density of an infinitesimal region. Rather, for the observable, which is the coarse-grained (macroscopic) phase-space density, we simply must not exceed the maximum of the fine-grained one.

Thus, the SN phase-space density must not increase during collapse, from its starting value of $\half g_{\nu}h^{-3}m_{\nu_s}^{4}$ (which is half the Pauli degeneracy limit), to its current maximum value of $\rho_{\nu_s}(r)[2\pi\sigma_{\nu_s}^2(r)]^{-1.5}$ (where we assume the velocity distribution is locally Gaussian everywhere with dispersion $\sigma_{\nu_s}$). This limit is called the Tremaine-Gunn limit (\citealt{tremaine79}; hereafter TG) and rearranging it in terms of the critical density for an $11~eV/c^2$ SN, where $1~eV/c^2=8.9\times10^{-67}\msun$ and $s^3c^3=9.18\times10^{-25}pc^3$ (where $s$ is obviously 1 second) gives us

\beq
\protect\label{eqn:tg}
\rho_{\nu_s,TG}(r)=3.15\times10^6\left({\sigma_{\nu_s}(r) \over c} \right)^3\msun pc^{-3}.
\eeq
The important thing to bear in mind here is that this is the exact value of the TG limit because it is calculated from the unique, derived SN VD exactly necessary for hydrostatic equilibrium and not by assuming some relation between the ICM and the neutrinos, as is often the case (\citealt{afb,natarajan08,gentile08}). Therefore, we have the TG limit as a function of radius, since we know the exact value of the VD from solving Eq~\ref{eqn:nueq2}.

 Let us note however that departures from Gaussianity in the velocity distribution could perhaps somewhat affect this physical upper limit on the SN density.

The consequence of the TG limit is that, therefore if the density of SNs required for cohesion of the cluster gas exceeds TG limit (in particular at the centre) then the MOND plus $11~eV/c^2$ SN hypothesis would be ruled out.

\section{Results}

In Figs~\ref{fig:c1}-\ref{fig:bul}  we present the densities, VDs and enclosed mass profiles for the $\numb$ groups and clusters. In the left hand panels we plot the SN density (solid line type), ICM density (dotted) and the TG limit of the cluster (dashed) against radius. In the middle panels we plot both the observed ICM (dotted) and the derived (from Eq~\ref{eqn:nueq2}) SN (solid) VD as functions of radius. In the right hand panels, we plot each cluster's enclosed Newtonian dynamical mass (dashed), MOND dynamical mass (dot-dashed), BCG total mass with unity $M/L$ (solid) and ICM mass (dotted).

There are a few salient features to observe: primarily, the SN VD (or temperature) is in most cases 20-50 per cent lower than the ICM VD. The simple explanation, alluded to earlier, is that the SN halos presumably formed at relatively high redshift, through only their mutual gravitation. On the other hand, the ICM fell from large distances, through the already present, deep potential well of the SNs and thus had greater potential energy to transfer to kinetic energy, although this still has to be demonstrated with numerical simulations. 

Taking a closer look at the SN densities and VDs, they have conspicuous kinks around 20-30~$kpc$. This radius, $r_{tg}$, is where the density reaches the TG limit, . If no phase space limit existed, the SN density would continue to increase towards the Pauli limit and the neutrino equation of state would begin to substantially change from non-degeneracy (large negative $\chi$) $P_{\nu_s}\propto \rho_{\nu_s}\sigma^2_{\nu_s}$ to the degenerate one (large positive $\chi$) $P_{\nu_s}\propto \rho^{5/3}_{\nu_s}$ (where the TG limit occurs at $\chi \sim 0.35$). However, since we must adhere to the TG limit, we impose it by making the density equal to the TG limit. The density and VD both increase towards the centre as they must conform to the duel constraints of remaining at the TG limit and satisfying hydrostatic equilibrium. Alternatively, if we had ignored the TG limit, the VD would have crashed towards the centre of the cluster; as less classical pressure would be required to prop up the SN halo. This is simply because gradually more Fermi pressure would be available. It is this behaviour that causes the kink since we force the density to be equal to the TG limit, the required VD for hydrostatic equilibrium must rise, whereas it was falling prior to this.

Since the SN halo of every group and cluster reaches the TG limit at the centre when there is still a discrepancy between the central SN density and the required total density for hydrostatic equilibrium of the gas, the BCG plays a crucial role in determining how satisfactory our $11~eV/c^2$ SN hypothesis is. If no BCG existed in any of our clusters, they would immediately fail.

We know the luminosity of most BCGs in the K-band which gives an excellent indication of mass (see e.g. \citealt{belldejong,conroy09,gastaldello07,humphrey06}) and we know a few in other bands (B,V,R) or not at all. Therefore, we must modify the $M/L_K$ (given in table~1) to exactly match $M_m$ at $r_{tg}$, which basically means the BCG is picking up all the ``slack'' left when the SN reaches the TG limit and can no longer account for the full dynamical mass. 

There are some points to bear in mind, firstly the highest $M/L_K$ used was 1.2 and the lowest 0.1. In the latter case, of a very low $M/L_K$, this can be due to the total luminosity not being enclosed by $r_{tg}$ which is often $\sim20-30~kpc$). In the case of $M/L_K \sim 1$, this can be considered a fitting parameter, since there is no freedom in the cluster to have a $M/L_K$ lower than the value used because the SN density reaches the TG limit at the centre of every cluster. Furthermore, the $M/L_K$ cannot be significantly larger or it would be in disagreement with the typical $M/L_K$ demonstrated by \cite{belldejong}. In the cases where $M/L_K < 0.8$ it need not be considered a fitting parameter, since there is freedom for the $M/L_K$ to be larger and simply for the SN to be lower. Nevertheless, the TG limit will still be reached in every cluster, the only difference will be that if we are underestimating the BCG mass then $r_{tg}$ will simply be lower for that particular cluster. The only reason we fix the density to be equal to the TG limit for all radii smaller than $r_{tg}$ is to highlight the maximum amount of luminosity the BCG could lose (e.g. if observations are incorrect) and still provide the central density required.

Generally, it is interesting to note that the SN halos needed to fit galaxy clusters here have a density slope similar to that of the ICM in the central parts, but becoming sharper at intermediate distance, which is accompanied by a relatively flat velocity dispersion. At the edges of the clusters, the ICM density becomes larger than the SN density (which falls to zero) and there is an apparent sharp decrease of the sterile neutrino velocity dispersion to zero also. This is merely a numerical artefact of the sterile neutrino density being set to zero at the edges of clusters, under the assumption of hydrostatic equilibrium. Since the sterile neutrino density will in reality fall to a very small number, but still greater than zero, the velocity dispersion will actually be isothermal, as we would expect.

\subsection{Individual systems}
Below we discuss some pertinent observations about individual groups or clusters.

\subsubsection{The bullet cluster}
\protect\label{sec:bul}
Analysis of the CMB strongly favours the hypothesis that non-baryonic DM exists in the Universe, and the bullet cluster (\citealt{clowe06}) compounds it (although \citealt{milgrom08} has suggested that the DM of clusters could be ultra cold and collisionless clumps of molecular gas which would satisfy the constraint of this particular cluster). As two giant galaxy clusters crashed into each other at incredible speed (\citealt{markevitch02}, but see also \citealt{am08} for why this might pose a problem for $\lcdm$), mutual ram pressure from the 2 ICMs dragged one another out of the clusters leaving only the galaxies and DM to pass through and emerge on the opposite sides on the sky. A weak-lensing reconstruction required two DM halos to overlay the positions of the galaxies with NFW parameters for the main and sub cluster respectively - $M_{200}=15.0, 1.5\times10^{14}\msun$, $r_{200}=2.1, 1.0~$Mpc and concentrations $c=1.94, 7.12$.

In this case, where we are directly given the Newtonian mass profile, we begin the procedure at Eq~\ref{eqn:mm} and follow the same steps. As with the other clusters, in Fig~\ref{fig:bul} we show that these two clusters both reach the TG limit smoothly at the centre. However, notice we must add a $3.5\times10^{11}\msun$ stellar mass to the sub cluster (labeled ``Bullet 2'') to cover the dynamical mass in the central $30~kpc$, which is very significant. The stellar mass quoted in \cite{clowe06} at the weak-lensing peak of the sub cluster is $(5.8\pm0.9)\times10^{11}\msun$ within 100~$kpc$, from I-band observations assuming a $M/L_I$ of 2. We need only add a trivial stellar mass of $1.0\times10^{11}\msun$ to the main cluster, which is interesting since one can see from the left hand panel of Fig~1 from \cite{clowe06} that there is no obvious BCG candidate associated with it. In fact, the two giant ellipticals of the main cluster highlighted by \cite{clowe06} are roughly 50~$kpc$ (to the northern one) and 75~$kpc$ (to the eastern one) from the lensing peak, but can easily offer the necessary stellar mass. On the other hand, the centre of the BCG of the sub cluster is only roughly 25~$kpc$ from the lensing peak and significant light is spilling over within 10~$kpc$. Furthermore, since BCGs usually dominate the stellar mass in the central 100~$kpc$ of clusters, the majority of the $5.8\pm0.9\times10^{11}\msun$ is likely associated with it, making the $3.5\times10^{11}\msun$ plausible. Notice also that the $3.5\times10^{11}\msun$ is not required within $10~kpc$, but rather by $30~kpc$. Within $10~kpc$ less than $1.5\times10^{11}\msun$ is required.

\subsubsection{RGH~80}
Another intriguing point about BCG masses is the one used for the group RGH~80, which has two equally massive central galaxies: NGC~5098a and NGC~5098b with $L_K=2.9\times10^{11}\lsun$ and $L_K=2.4\times10^{11}\lsun$ respectively. In our previous paper (\citealt{afb}) we were only interested in the central $100~kpc$, therefore, were able to combine the two galaxy luminosities together. However, only one of the galaxies is at the very centre (see Fig~9 of \citealt{mahdavi05} or Fig~1 of \citealt{randall09}), and the other has a projected separation of around $50~kpc$ and possibly a considerable line of sight distance. Before removing the second galaxy, the TG limit was considerably larger than the maximum central density, but after discounting it, the TG limit is reached at $20~kpc$. 
\subsubsection{Clusters with no BCG luminosity: A~1689 and A~2390}

We found no galaxy luminosities for the two clusters A~1689 and A~2390, but they require BCGs with total masses of at least $2.0$ and $18.5\times10^{11}\msun$ respectively, which is a prediction of this model. Interestingly, A~1689 has been used extensively to argue against CDM by \cite{broadhurst08} because the observed NFW concentration parameter is considerably larger than that expected from cosmological simulations. We find the equilibrium model with $11~eV/c^2$ SNs nicely reaches the TG limit at $20~kpc$, even though it has one of the highest central SN densities of all our sample.

\subsubsection{Lynx}
The two clusters in the Lynx field are in the process of merging (like the bullet cluster) and thus the mass profiles are potentially overlapping. Nevertheless, their offset seems to be sufficiently large to get a reasonable estimate from weak lensing and this has also been sanity checked with X-ray hydrodynamics (\citealt{jee06}). Both the larger cluster and smaller cluster are fitted with the same NFW profile: $M_{200}=2.0\times10^{14}\msun$, $r_{200}=0.75~$Mpc and concentration $c=4$. The galaxy luminosities within 500~$kpc$ of each of the two lensing peaks are 1.5 and 0.8$\times10^{12}\lsun$ in the B-band. We require $6.0\times10^{11}\msun$ for each BCG, which is easy affordable by the luminosity of both clusters because typical B-band $M/L$s  can range between 5 and 10.

\subsubsection{Coma}
The historical significance of the Coma cluster with respect to the dark matter problem is probably far more significant than its scientific significance in this present case, but we include it out of curiosity. It was originally analysed in MOND by \cite{the88} and was actually concluded to be more or less consistent with MOND, although the galaxies required radially biased orbits and the acceleration constant, $a_o$ had to be increased by at least a factor of 2 from the one used here. The problem with Coma is that it is not particularly relaxed and also its sphericity is questionable. For example, \cite{neumann03} showed that there is ongoing merging, which makes measurements of the central mass profile uncertain.  Therefore, the only way to get a decent estimate of the dark halo of Coma is to use weak lensing (like we have done with the bullet cluster), but even then the assumption of spherical symmetry is dubious. The best study of Coma was performed by \cite{gavazzi09}, but unfortunately found only a rather speculative NFW profile of $M_{200} = 5.1^{+4.3}_{-2.1} \times10^{14}\msun$ and $r_{200} = 1.8^{+0.6}_{-0.3}Mpc$ with no prior on the concentration parameter (found to be $c_{200} = 5.0^{+3.2}_{-2.5}$). With a prior on the concentration parameter (set to be $c_{200} = 3.5^{+1.1}_{-0.9}$) they found $M_{200} = 9.7^{+6.1}_{-3.5} \times10^{14}$ and $r_{200} = 2.2^{+0.3}_{-0.2}Mpc$. Finally, we added an estimate from \cite{kubo07}: $M_{200} = 27.0\pm 8.0 \times10^{14}$, $r_{200} = 2.9 \pm 0.3~Mpc$ and $c_{200} = 3.8^{+13.2}_{-1.8}$.  The mass of the central galaxy would need to be $1.5, 1.0$ and $2.0\times10^{11}\msun$ respectively for the three cases. The K-band luminosity of the central galaxy in Coma is $10^{12}\lsun$, so there is plenty galactic mass to supply what is needed.

\subsubsection{RXJ~1416}
We included one fossil group RXJ~1416 in the sample since it has a very minor galaxy component (aside from the BCG). \cite{khosroshahi06} fitted an NFW profile to X-ray data from Chandra and XMMN with parameters $M_{200} = 3.1 \times10^{14}\msun$, $r_{200} = 1.2~Mpc$ and $c_{200} = 11.2\pm 4.5$ and we used it to infer the necessary $M/L$ of the BCG which has luminosity $L_R=7\times10^{11}\lsun$. The $M/L_R$ need only be 1.2, whereas 5 is a typical value for an old stellar population in that band. So here again, as with several of the other massive systems, the full luminosity of the BCG is not required although the SN halo reaches the TG limit at the centre. This might be because the BCG is extended and only a fraction of the light is enclosed within 20 or 30$~kpc$ (where the SN halo reaches the TG limit). Alternatively, if we used the full mass of the BCG, the SN density would simply reach the TG limit at a smaller radius.

\subsubsection{NGC~4125 and NGC~6482}
\protect\label{sec:n4125}
These two X-ray bright early type galaxies were taken from the sample of \cite{humphrey06}. Interestingly from the point of view of SNs in galaxies (which could disrupt the good fits to rotation curves i.e. the basis of MOND) their dynamical masses are comfortably covered by the BCG mass everywhere. From their density figures you can see the SNs are at the TG limit everywhere, but the SNs make virtually no impact on their mass profiles. In fact, if the SN density for NGC~4125 was $8 \times 10^{-4}\msun pc^{-3}$ all the way to the centre from 10~$kpc$, the enclosed mass of SNs would be $3.4\times10^9\msun$ in comparison to roughly $1.0\times10^{11}\msun$ for the BCG (at that radius), which is a factor of 30. Therefore, there is no reason to believe that SNs will influence the internal dynamics of individual galaxies in a meaningful way.

\subsubsection{NGC~720 and NGC~1550}
\protect\label{sec:n720}
These are the two most problematic groups. \cite{buote94,buote96,buote02} observed that the twisting of X-ray isophotes around the elliptical galaxy NGC~720 compared to the intrinsic ellipticity of the galaxy (which outweighs the gas by more than two orders of magnitude) could only be compatible with the presence of DM at least four times the galaxy mass (by the edge of the galaxy). This SN halo would have an ellipticity distinct to that of the galaxy and the gas would trace the superimposed potential, hence generating the twisting isophotes. The luminosity of NGC~720 is $L_K=1.7\times10^{11}\lsun$ and requires $M/L_K=1.2$ which is the high end of the scale for the Kroupa IMF of an old stellar population. The SNs have reached the TG limit by 30$~kpc$, so cannot contribute any more mass. The problem is that NGC~720 has a relatively young stellar population, although there is a age gradient from $\sim 12~Gyr$ in the centre to $\sim 3~Gyr$ by $\sim 1~kpc$ (\citealt{humphrey06}). This leads to significant systematic uncertainty in constraining the $M/L_K$, which is given as $0.54\pm0.11$ and $0.35\pm0.07$ for the Salpeter and Kroupa IMF respectively. A recent re-evaluation with data of superior resolution (D. Buote private communication) puts the Kroupa value at $0.49\pm0.18$ (meaning the Salpeter value will be somewhat larger), but this still falls well short of the necessary galactic mass. This could be a serious problem if the low $M/L_K$ could be confirmed, but for now it is a prediction of this model that the true $M/L_K$ (when the correct IMF, age distribution and merger history are taken into account) will be close to 1.2.

NGC~720 is not the only system that is very sensitive to the observations. NGC~1550 was studied by \cite{gastaldello07} with both Chandra and XMM-Newton and both sets of temperature data are plotted in their Fig~3. Later it was observed again by \cite{kawaharada09} using XMM-Newton, in a study which suggested there was evidence for a recent merger. If we use either set of XMM-Newton data points, the $M/L_K$ would need to be at least 1.6, whereas if we only subscribe to the Chandra data, the $M/L_K$ need only be 1.2. Follow up observations could provide a very strong test of the model, but Chandra's greater spatial resolution makes it the more reliable data set.

\subsection{Sterile neutrino mass and the $\mu$-function}
\cite{angus09} demonstrated that $11~eV/c^2$ SNs were required to fit the first three acoustic peaks of the CMB and that 2 species of $5.5~eV/c^2$ SNs were totally inadequate (as were three $2.2~eV/c^2$ ordinary neutrinos). There was also very little freedom beyond 5 per cent of $11~eV/c^2$, without introducing unjustified free parameters.

For further instruction to the importance of the SN mass to be $11~eV/c^2$, we calculated the $M/L_K$ necessary to produce an equilibrium SN distribution in the group ESO-306 (which is by no means the most constraining) in two different scenarios. Firstly, trying a $9.5~eV/c^2$ SN and secondly keeping an $11~eV/c^2$ SN mass but using the standard $\mu$-function of MOND: where instead of using Eq~\ref{eqn:fb} we try $\mu(g/a_o)={g/a_o \over \sqrt{1+(g/a_o)^2}}$. Whereas using the simple $\mu$-function and an $11~eV/c^2$ SN requires $M/L_K=1$, the $9.5~eV/c^2$ SN requires $M/L_K=1.5$ and the standard $\mu$-function would need 1.4.
 Given that these sorts of high $M/L_K$ would be the rule, rather than the exception, they seem incompatible with stellar population synthesis models like those of \cite{belldejong}.

This evidence in favour of the simple $\mu$-function is not an isolated case. In \cite{fb05} and \cite{mcgaugh08}, the simple function was prefered from a fit to the Milky Way's rotation curve (a high surface brightness galaxy) as was the case in the large sample of high surface brightness galaxies carried out by \cite{sandnoord}. This preference is also being found by an ongoing study of ultra high resolution rotation curves observed by the THINGS (\citealt{things1}) collaboration (G. Gentile, private communication). The main problem with the simple function is that it cannot be used all the way to the strong gravity regime. In particular, it would produce too high a modification in the inner Solar System, which is excluded, for example, by measures of the perihelion precession of Mercury. The solution is to use a $\mu$-function that would rapidly interpolate between the simple and standard $\mu$ for very large values of the gravitational acceleration, but this is not applicable here.

\subsection{Common central phase-space density}

It is an encouraging result that the $11~eV/c^2$ SNs have unique and continuous equilibrium models for such a large dynamic range of cluster properties, but what is remarkable is that at the centre of every cluster, the TG limit is reached. 
For each given cluster, there is a specific, maximum density profile that can exist in equilibrium: a good example being NGC~5129 (in Fig~\ref{fig:c3}), which must have the maximum allowed density from $40~kpc$ for the SN halo to both be in hydrostatic equilibrium and provide the correct dynamical mass as measured by the ICM properties.

Accordingly, it is the TG limit and the degenerate properties of the SNs (since they are fermions) that sets the dynamical properties of all clusters. No relation, even remotely like this exists if the cluster DM is cold or non-fermionic.

\subsection{Sterile neutrinos inside galaxies}
\subsubsection{Dynamics}
It is apparent from the examples of NGC~4125 and NGC~6482 (see \S\ref{sec:n4125}) that SNs will mostly not affect the dynamics of galaxies. \cite{gentile08} have shown that a constant density of $10^{-5}\msun pc^{-3}$ was allowed in the MOND fits to the rotation curves of Ursa Major galaxies. This magnitude of SN density is typically found within $1~Mpc$ in the very massive clusters and within $100-300~kpc$ from the centre of groups. At the centre of clusters, the situation is obviously different, but stable spiral galaxies are never present there since tides would rip them apart before they fall to the centre. Therefore, this will not disturb the MOND Tully-Fisher relation since field spirals should be far from any SN halos.

 However, SNs are required to exist in the centres of some very massive ellipticals. This is clear from the right hand panels of the enclosed mass profiles of the clusters A~478 (Fig~\ref{fig:c1}), A~907  and others. In these figures one can see the mass of the BCG (the solid line) is smaller than the MOND enclosed mass. Thus, a considerable mass of SNs is required inside the limits of the stellar orbits, but always significantly less than the mass in stars.

\subsubsection{Lensing studies of individual galaxies in MOND}
In its original form, MOND should be able to explain all galaxy dynamics without dark matter. Although rotation curves have always yielded excellent results, the data for lensing studies (beginning with \citealt{zhao06}) of individual galaxies has not been as promising. For instance \cite{tian09} show that the weak-lensing of single, isolated galaxies is perfectly compatible with MOND up to a particular galaxy luminosity (from $L_r=0.1 - 8.0\times10^{10}\lsun$). Thereafter, the lensing data implies the need for DM, which is exactly as we might expect if these more luminous galaxies are embedded in a low density but extended SN halo (akin to those in the \citealt{humphrey06} sample, see \S\ref{sec:n4125}). The same is true for the study by \cite{ferreras08} whereby the integrated mass along the line of sight is entangled with the stellar mass. This makes lensing studies of individual early-type galaxies in MOND significantly inferior to those using ICM, or even globular clusters (\citealt{richtler08}), planetary nebulae (\citealt{romanowsky03,milgrom03,douglas07,napolitano09}) and satellite galaxies (\citealt{klypin09,aftcz}).

\section{Discussion and Conclusion}

In this paper we have shown explicitly how to calculate the equilibrium configurations of neutrinos in MONDian galaxy clusters. The density of sterile neutrinos is fixed by the properties of the ICM (or the observed lensing map), but derivation of the sterile neutrino velocity dispersion allows a specific density profile to exist in hydrostatic equilibrium. We have presented the detailed properties of $\numb$ typical galaxy groups and clusters over a wide range of masses and temperatures (and even redshift vis-a-vis the bullet cluster, the lynx cluster and A~1689) and have shown that not only can we elucidate velocity dispersion profiles that allow the sterile neutrinos to exist in hydrostatic equilibrium, but also that the Tremaine-Gunn limit sets the central density.

It would appear to be a very strong coincidence that by doing little more than fixing the mass of a sterile neutrino to be $11~eV/c^2$, we can serendipitously, explain the formation of the acoustic peaks in the CMB and specify the {\it exact} properties of systems that require DM in MOND. In particular, regardless of cluster mass, the velocity dispersion of sterile neutrinos necessary to impose hydrostatic equilibrium allows the density to reach its maximum (which is a function of both sterile neutrino particle mass, velocity dispersion and cluster mass indirectly) at the centre of the cluster. The only stipulation is whether these equilibrium configurations are stable, which should be the next check.

Out of the $\numb$ systems, there are two which need to be monitored. NGC~720 requires a K-band mass-to-light ratio of 1.2, which is possible for an old stellar population, but the current best population synthesis model suggests $0.49\pm0.18$. Estimates of stellar masses are notoriously fraught with difficulties (\citealt{conroy09}), but this is a potential problem. NGC~1550 (discussed alongside NGC~720 in \S\ref{sec:n720}) can only reproduce the temperature observed from Chandra; if the XMM-Newton data are used, the K-band mass-to-light ratio of the central galaxy is too high. From this study, it is clear that the strongest tests of the $11~eV/c^2$ sterile neutrino model comes not from the rich clusters, but rather from smaller groups or individual galaxies with bright X-ray halos. 
 
\begin{figure*}
\def\subfigtopskip{0pt} 
\def\subfigbottomskip{4pt}
\def\subfigcapskip{1pt}
\centering
\begin{tabular}{c}
\subfigure{
\includegraphics[angle=0,width=16.0cm]{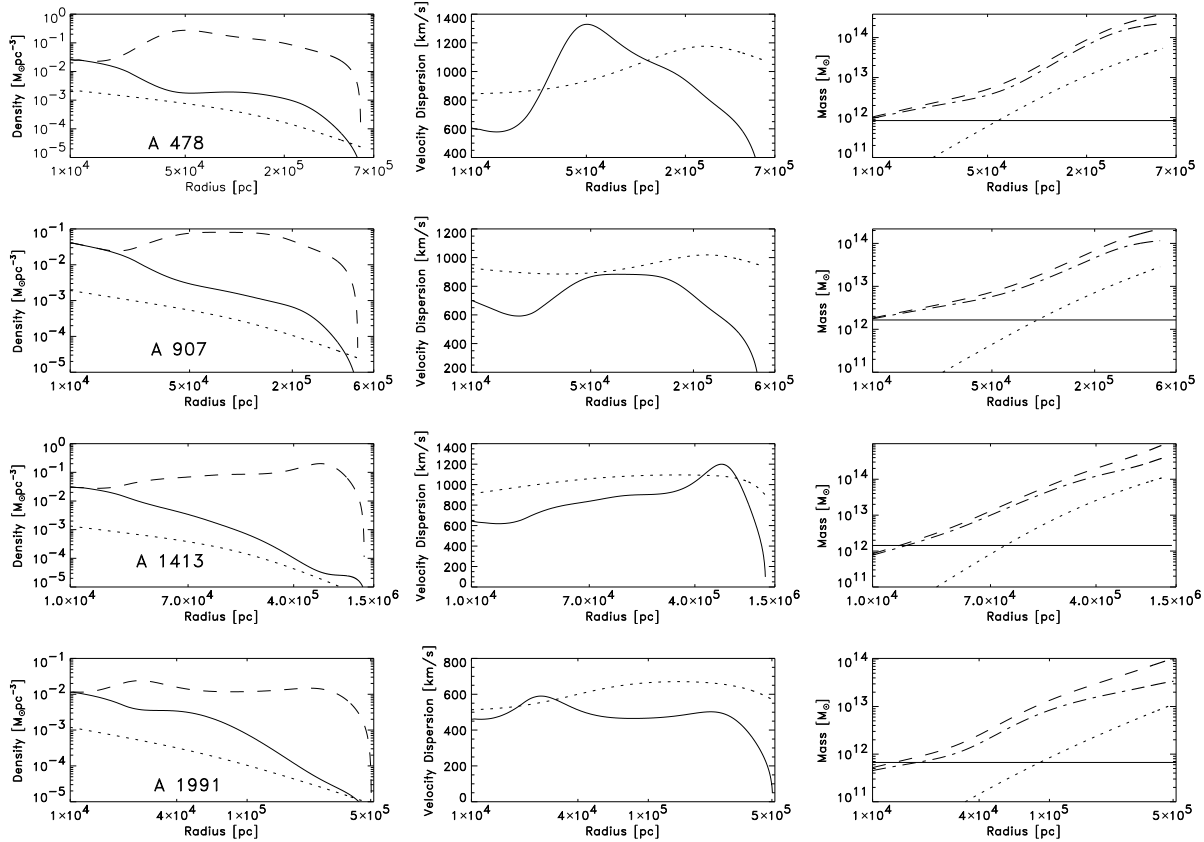}
}\\
\end{tabular}

\caption{As per Figs~\ref{fig:c2} to \ref{fig:bul}, each row refers to a specific cluster, whereas the columns are properties of that cluster. In the left hand panels we plot density against radius for sterile neutrinos (solid), intracluster medium (dotted) and Tremaine-Gunn limit (dashed). In the central panels we plot velocity dispersion against radius for sterile neutrinos (solid) and intracluster medium (dotted). In the right hand panels we plot enclosed mass against radius for the Newtonian Dynamical mass (dashed), MOND dynamical mass (dot-dashed), intracluster medium (dotted) and the total mass of the brightest cluster galaxy (solid). Each cluster's designation is inscribed on the left hand panel.}
\label{fig:c1}
\end{figure*}

\begin{figure*}
\def\subfigtopskip{0pt} 
\def\subfigbottomskip{4pt}
\def\subfigcapskip{1pt}
\centering
\begin{tabular}{c}
\subfigure{
\includegraphics[angle=0,width=16.0cm]{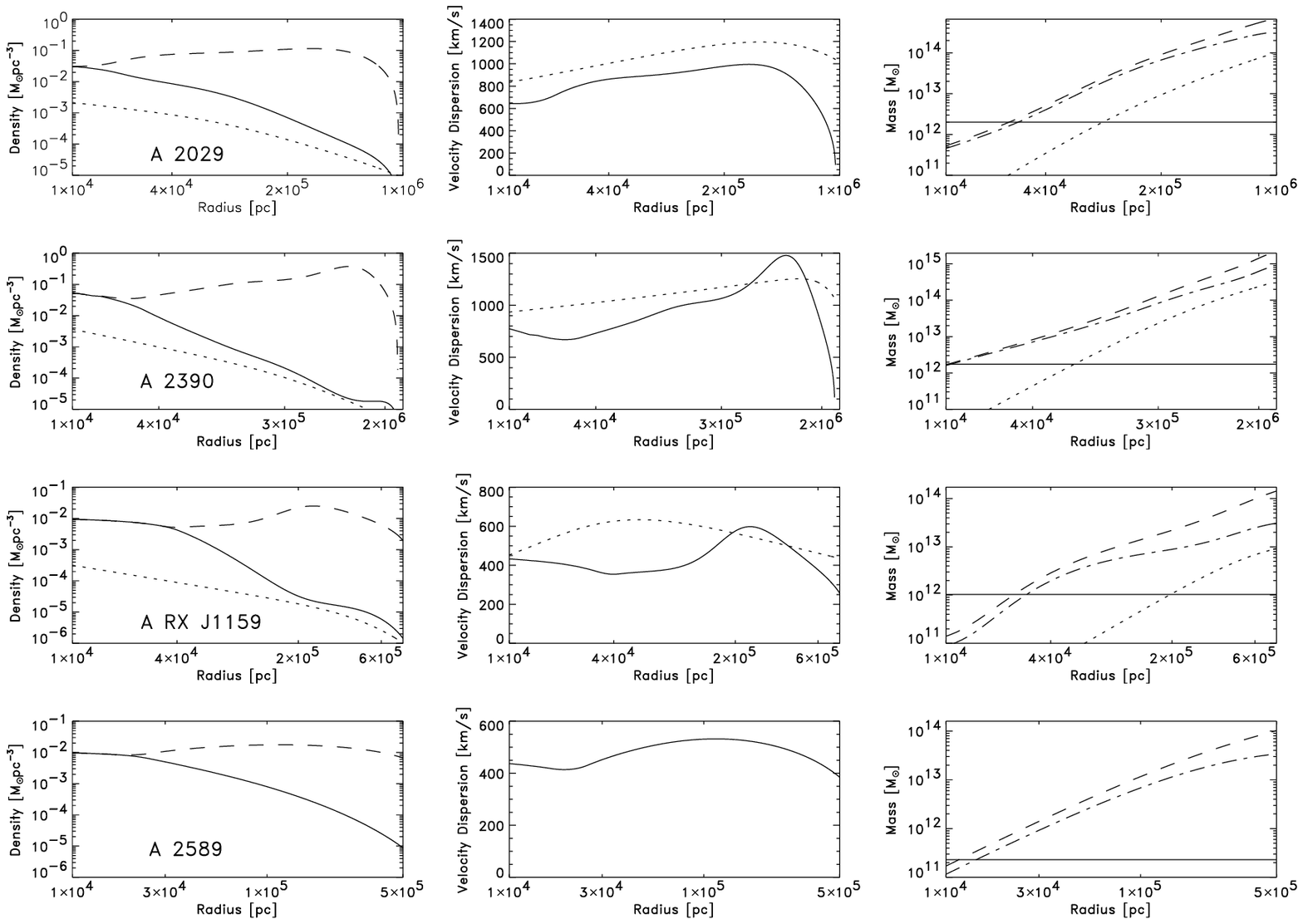}
}\\
\end{tabular}

\caption{As per Fig~\ref{fig:c1}.}
\label{fig:c2}
\end{figure*}

\begin{figure*}
\def\subfigtopskip{0pt} 
\def\subfigbottomskip{4pt}
\def\subfigcapskip{1pt}
\centering
\begin{tabular}{c}
\subfigure{
\includegraphics[angle=0,width=16.0cm]{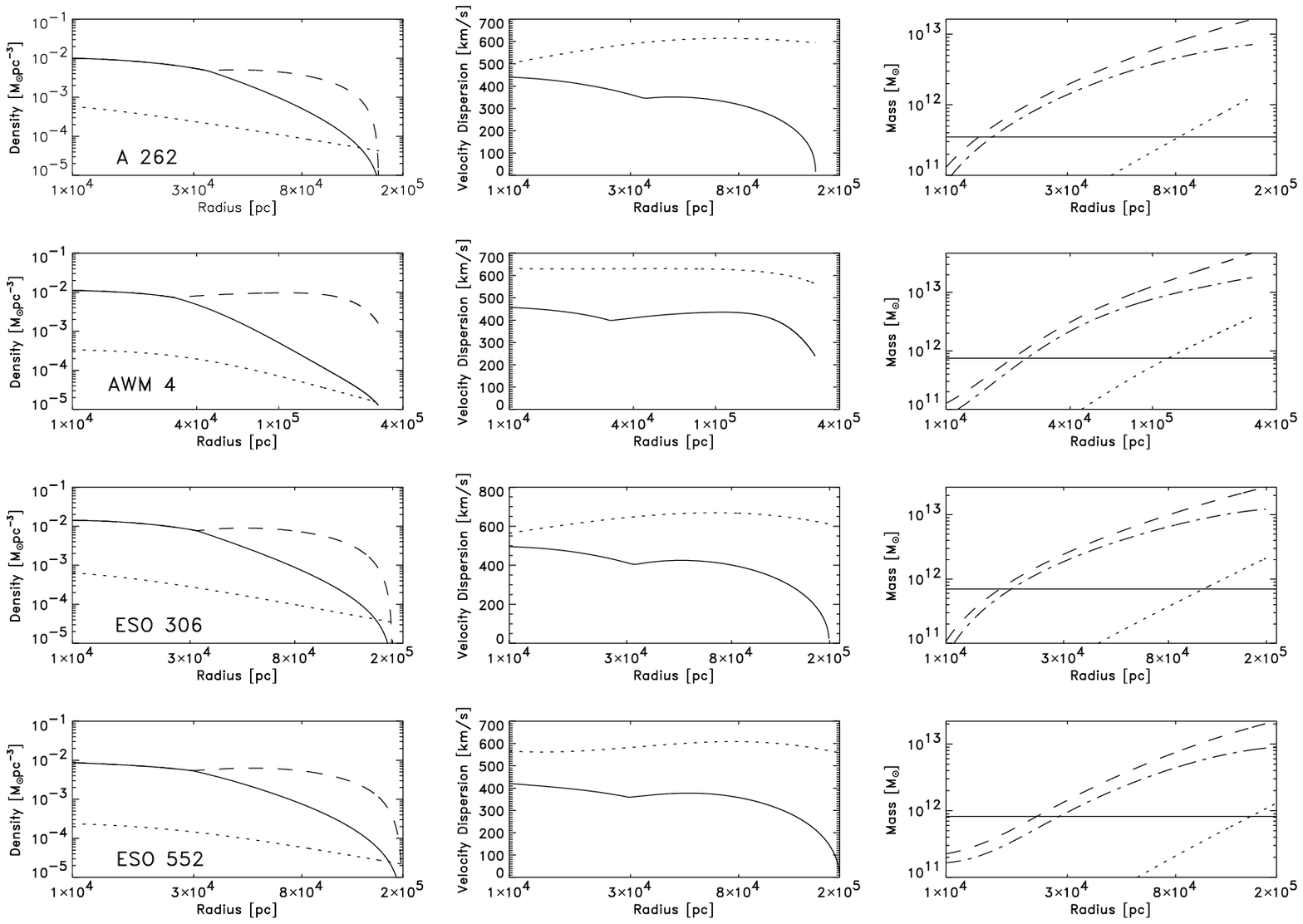}
}\\
\subfigure{
\includegraphics[angle=0,width=16.0cm]{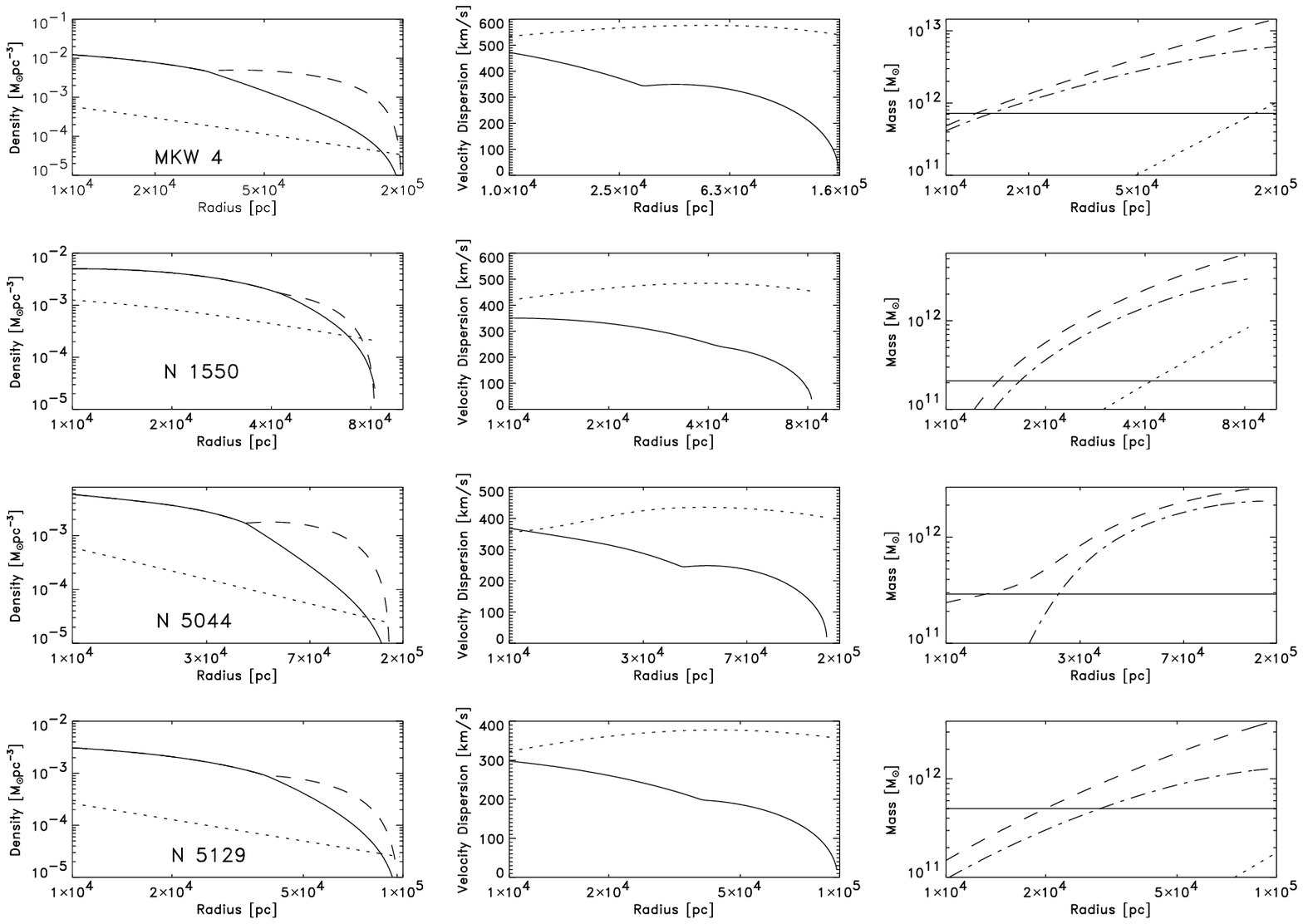}
}
\\
\end{tabular}
\caption{As per Fig~\ref{fig:c1}.}
\label{fig:c3}
\end{figure*}

\begin{figure*}
\def\subfigtopskip{0pt} 
\def\subfigbottomskip{4pt}
\def\subfigcapskip{1pt}
\centering
\begin{tabular}{c}
\subfigure{
\includegraphics[angle=0,width=16.0cm]{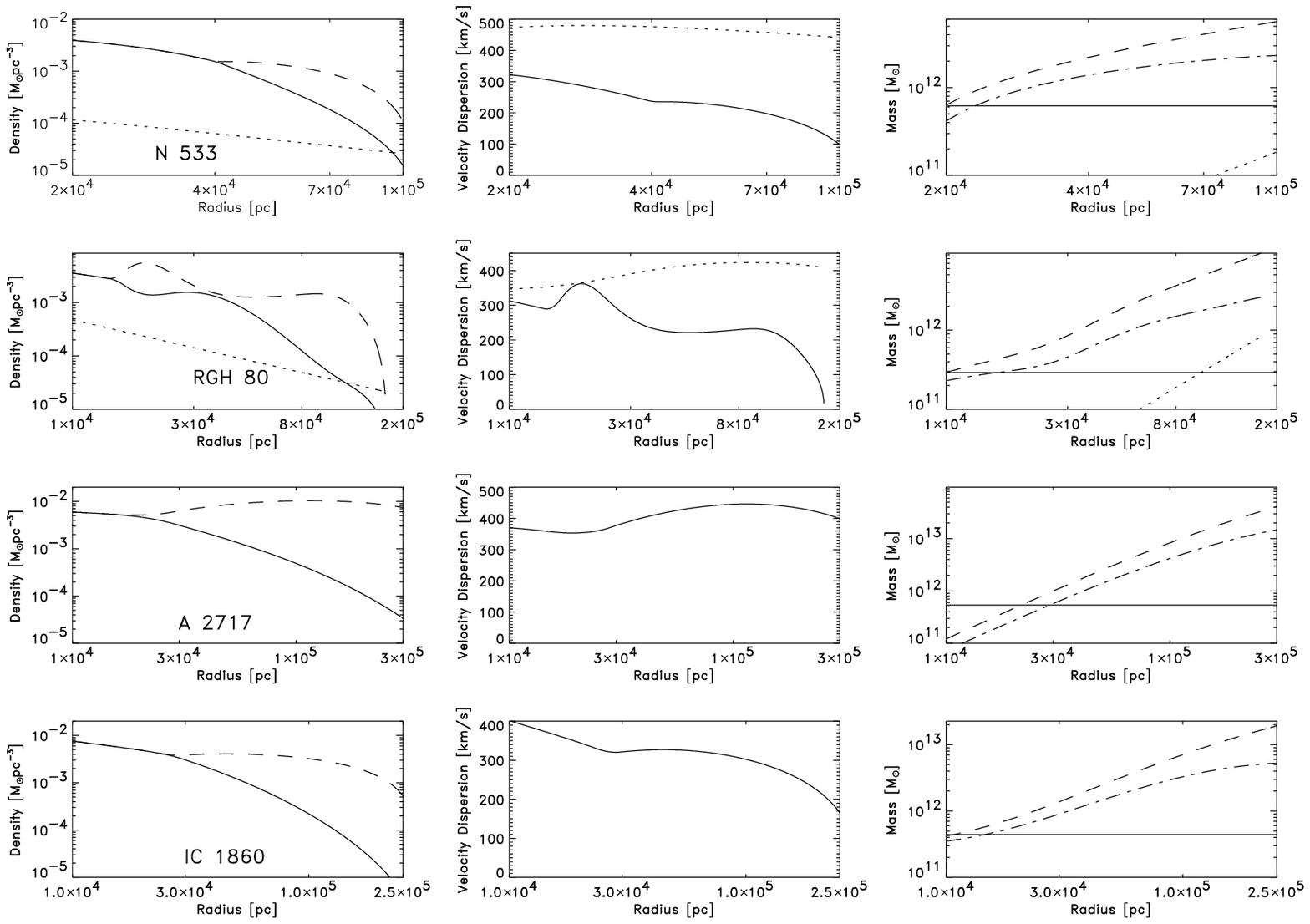}
}\\
\subfigure{
\includegraphics[angle=0,width=16.0cm]{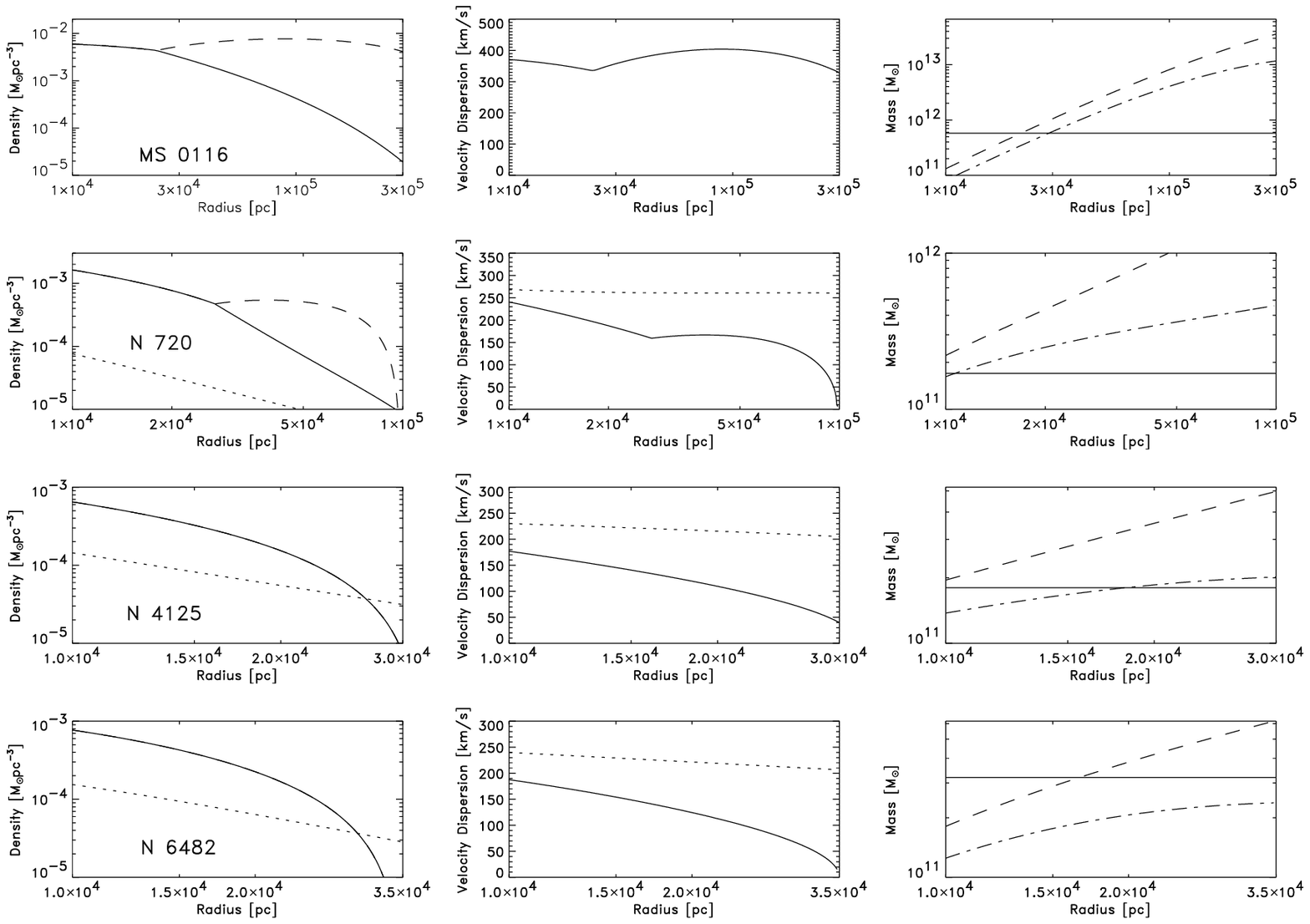}
}
\\
\end{tabular}
\caption{As per Fig~\ref{fig:c1} except certain lines corresponding to the intra-cluster medium are absent because the sterile neutrino densities were deduced from weak or strong lensing analyses.}
\label{fig:chum}
\end{figure*}

\begin{figure*}
\def\subfigtopskip{0pt} 
\def\subfigbottomskip{4pt}
\def\subfigcapskip{1pt}
\centering
\begin{tabular}{c}
\subfigure{
\includegraphics[angle=0,width=16.0cm]{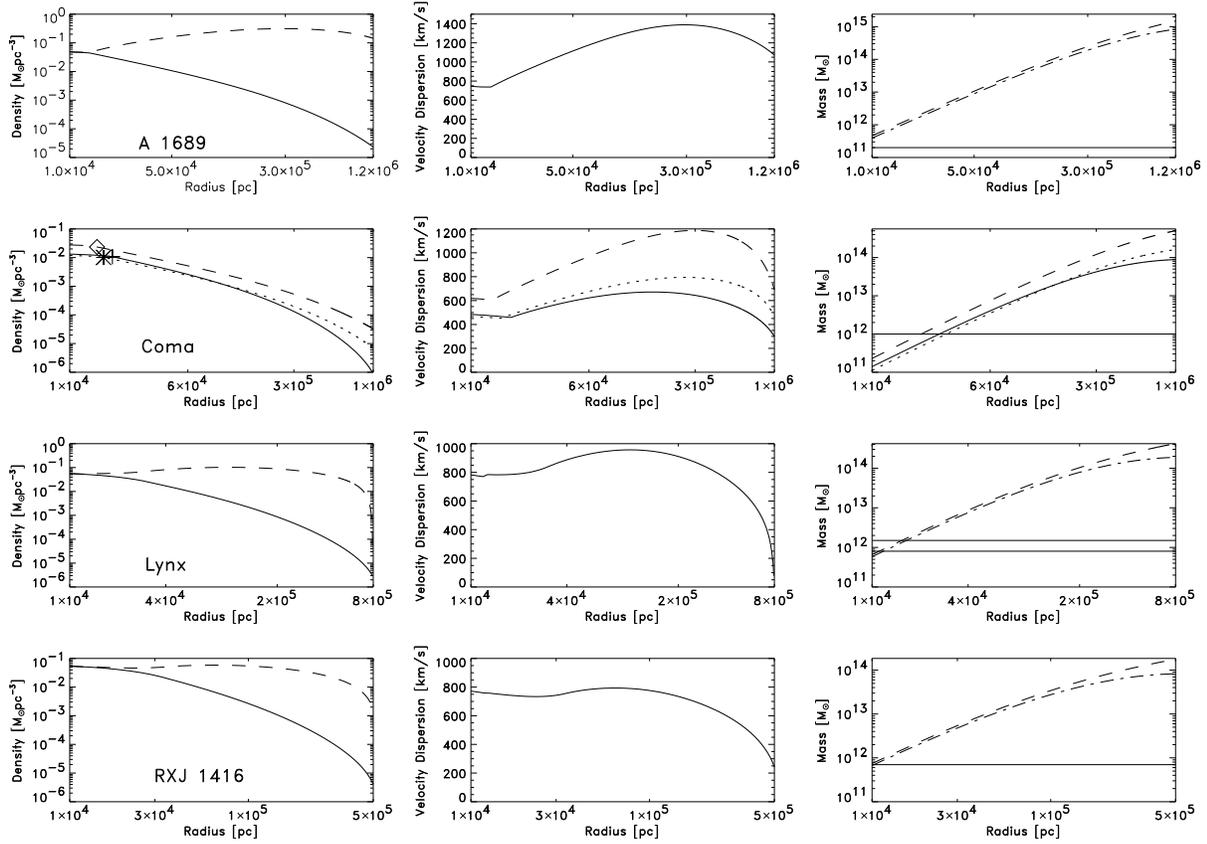}
}\\
\end{tabular}
\vspace{0 mm}
\caption{As per Figs~\ref{fig:c1}-\ref{fig:chum} except that they are for A~1689 (Halkola et al. 2006), the Coma cluster, the two clusters that comprise the Lynx cluster (Jee et al. 2006) and the fossil group RXJ~1416 (Khosroshahi et al. 2006), from which the masses are taken in NFW form. For the Coma cluster we use 3 different measurements for the NFW mass profile as discussed in detail in \S\ref{sec:bul}. To avoid crowding the Coma density plot, we only plot the sterile neutrino density for NFW profile 1 (solid), 2 (dotted) and 3 (dashed) and not the Tremaine-Gunn limit, which is instead marked by an asterisk, plus sign and diamond respectively where the sterile neutrino density reaches it.}
\label{fig:coma}
\end{figure*}

\begin{figure*}
\def\subfigtopskip{0pt} 
\def\subfigbottomskip{4pt}
\def\subfigcapskip{1pt}
\centering
\begin{tabular}{c}
\subfigure{
\includegraphics[angle=0,width=16.0cm]{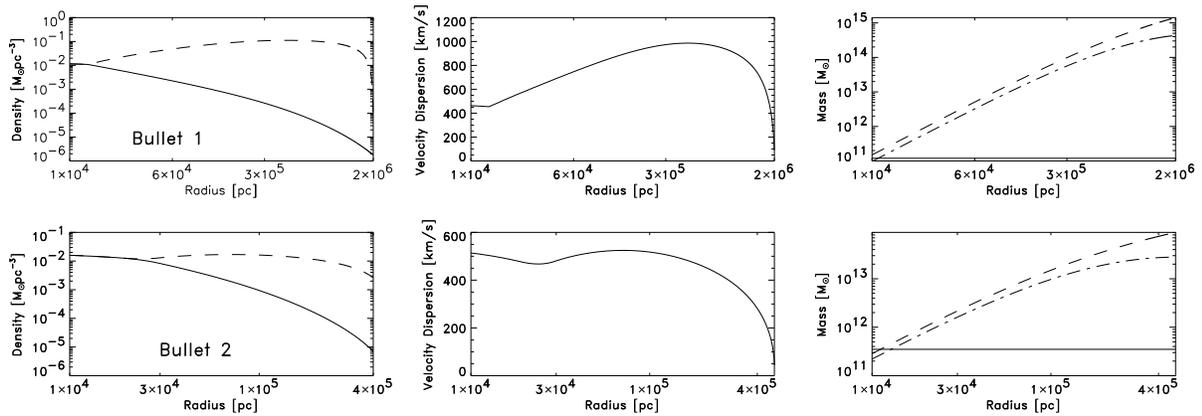}
}\\
\end{tabular}
\vspace{-50 mm}
\caption{As per Figs~\ref{fig:c1}-\ref{fig:coma} except that they are for the two clusters that comprise the bullet cluster (\citealt{clowe06}), from which the masses of the two clusters are taken in NFW form. This case is discussed in detail in \S\ref{sec:bul}}
\label{fig:bul}
\end{figure*}

In a similar sense to how the NFW density profile deduced from N-body simulations of CDM structure formation have been shown to be inadequate descriptions of some galaxy and galaxy cluster DM halos (\citealt{broadhurst08,deblok98,gentile04}) we suggest that our highly regular density and velocity dispersion profiles shown in Figs~\ref{fig:c1}-\ref{fig:bul} must be used to judge the consistency of simulated structure growth in MOND cosmological simulations. Presumably, it is guaranteed that the objects that condense out of the background will reach the Tremaine-Gunn limit at the centre (which makes it so alluring that all the clusters studied here do), but it is not certain that the phase space densities will fall with radius in the manner shown here, nor that the chemical potential will have the correct values.

Given that we fit the CMB and have these interesting results for clusters of galaxies, another crucial test of this model (after checking stability) will be to see if MOND N-body cosmological simulations, with $11~eV/c^2$ sterile neutrinos can form structures resembling those shown here and match the linear matter power spectrum.  It is worth pointing out that \cite{skordis06} computed the matter power spectrum with MOND-like gravity and three $2.75~eV/c^2$ active neutrinos (the $a_o$ used was $3.5\times$ larger than the one used typically to fit rotations curves and used here) and showed the extra matter density from the neutrinos ($\Omega_{\nu}=0.17$ compared to $\Omega_{b}=0.05$) coupled with the MOND gravity could come relatively close to providing a good match to the matter power spectrum measured by the SDSS (\citealt{tegmark04}). The higher matter density provided by the SNs ($\Omega_{\nu_s}=0.225$ and $\Omega_{b}=0.047$) traded off with the smaller, more standard $a_o$, could very well provide a superior fit to the matter power spectrum. To resolve this we need cosmological numerical simulations, which are still in their infancy (see \citealt{llinares08}).

Relativistic MOND theories beginning with TeVeS (\citealt{bekenstein04}, also see the review by \citealt{skordis09}) and BSTV (\citealt{sanders05}), which led to new ideas like Generalised Einstein-Aether (see e.g., \citealt{zlosnik07,zlosnik08}) are still highly complex and only address the galactic dark matter problem (not the cluster, or cosmological dark matter problem; nor the dark energy problem). Therefore, one caveat we would add to our conclusions is that this cosmological model still requires dark energy in the same coincidental amount as $\lcdm$. However, as elaborated upon in \cite{milgrom99} and \cite{milgrom08a}, MOND and dark energy must be two sides of the same coin that leads seamlessly to $2\pi a_o \approx c(\Lambda /3)^{1/2}$. There is some progress in this direction (\citealt{fuzfa07,bruneton09,li09,blanchet09}) but it has not yet been convincingly or efficiently shown.

\begin{table}
% use packages: array
\begin{tabular}{lll} 
Cluster & $L_K$ ($10^{11}\lsun$) & Min $M/L_K$ \\ 
A~262 & 4.1 & 1.0 \\ 
AWM~4 & 7.5 & 0.3 \\ 
ESO~306 & 7.0 & 1.0 \\ 
ESO~552 & 8.2 & 0.3 \\ 
MKW~4 & 7.2 & 1.1 \\ 
NGC~1550 & 2.1 & 1.2 \\
NGC~5044 & 2.9 & 1.1 \\ 
NGC~5129 & 5.0 & 0.6 \\ 
NGC~533 & 1.2 & 1.0 \\ 
RGH~80 & 2.9 & 0.8 \\
\\
A~478 & 8.4 & 1.2 \\
A~907 & 16.5 & 1.1\\
A~1413 & 18.3 & 0.4 \\
A~1991 & 6.9 & 0.65 \\
A~2029 & 20.1 & 0.2 \\
A~2390 & $18.5^*$ & ... \\
RXJ~1159 & 10.3 & 0.4 \\
\\
NGC~720 & 1.7 & 1.2 \\
NGC~4125 & 1.8 & 0.9 \\
NGC~6482 & 3.2 & 1.0 \\
\\
A~2589 & 2.3 (V) & 0.65 \\ 
\\
A~2717 & 5.4 & 0.2 \\ 
IC~1860 & 4.4 & 1.0 \\ 
MS~0116 & 5.8 & 0.25 \\ 
A~1689 & $2.0^*$ & ... \\
Coma & $\sim$10.0 & 0.15/0.1/0.2\\ 
Lynx & 15, 8 (B) & 0.4, 0.75   \\ 
RXJ~1416 & 7.0 (R) & 1.2 \\ 
\\
Bullet 1 & $1.0^*$ & ... \\ 
Bullet 2 & $3.5^*$ & ... \\ 
\end{tabular}
\caption{Here we list the K-band luminosities of our BCGs along with the minimum K-band mass-to-light ratio required to fit the dynamical mass. The clusters with a ($^*$) lack information about the BCG, so in the luminosity column, we have entered the required BCG luminosity with unity $M/L$. The clusters with a luminosity followed by (V),(B) or (R) have their luminosity measured in that band, and not the K-band. The Coma cluster has 3 separate mass profiles, so the 3 M/Ls refer to the profiles in the order they are taken in Fig~\ref{fig:coma}. The clusters are separated into samples: the top set are from the Gastaldello et al. (2007) sample; then the Vikhlinin et al. (2006) sample; Humphrey et al. 2006; Zappacosta et al. (2006); miscellaneous NFW fits; the bullet cluster.}
\end{table}

\section{acknowledgments} GWA's research is supported by the University of Torino and Regione Piemonte. Partial support from the INFN grant PD51 is also gratefully acknowledged. Support from the PRIN2006 grant ``Co\-stituenti fondamentali dell'Universo'' of the Italian Ministry of University and Scientific Research is gratefully acknowledged by AD. BF acknowledges the financial support of the Alexander von Humboldt foundation. We thank both David A. Buote and Fabio Gastaldello for providing the ICM temperature and density data in electronic format.

%\bibliography{fgc}

\end{document}